\documentclass[conference]{IEEEtran}
\usepackage{amsmath}
\usepackage{multirow}
\usepackage{setspace}
\usepackage{color,subfigure}
\usepackage{mathrsfs}
\usepackage{graphics}
\usepackage{epsfig}
\usepackage{url}
\usepackage[sort]{cite}
\usepackage[usenames,dvipsnames,table]{xcolor}
\usepackage[table]{xcolor}
\usepackage{booktabs}
\usepackage[T1]{fontenc}
\usepackage{tabularx}
\usepackage{blindtext}
\usepackage{graphicx}
\usepackage{ifpdf}
\usepackage[cmintegrals]{newtxmath}
\usepackage{bm}

\newcommand{\subcaption}[1]{\centerline{{\scriptsize
  #1}}\vspace{10pt}}
  \newlength{\minipagewidth}
\newlength{\figurewidthFour}

\def\BibTeX{{\rm B\kern-.05em{\sc i\kern-.025em b}\kern-.08em
		T\kern-.1667em\lower.7ex\hbox{E}\kern-.125emX}}

\begin{document}

\title{Speeding-up Age Estimation in Intelligent Demographics System via Network Optimization
\thanks{Identify applicable funding agency here. If none, delete this.}
}

\author{\IEEEauthorblockN{Zhenzhen Hu\IEEEauthorrefmark{1}\IEEEauthorrefmark{2},  Peng Sun\IEEEauthorrefmark{2}, Yonggang Wen\IEEEauthorrefmark{2}} 
	
	\IEEEauthorblockA{\IEEEauthorrefmark{1}School of Computer and Information, Hefei University of Technology, Hefei, China} 
	\IEEEauthorblockA{\IEEEauthorrefmark{2}School of Computer Science and Engineering, Nanyang Technological University, Singapore} \\
    huzhen.ice@gmail.com, \{sunp0003, ygwen\}@ntu.edu.sg
}

\maketitle

\begin{abstract}
Age estimation is a difficult task which requires the automatic detection and interpretation of facial features. Recently, Convolutional Neural Networks (CNNs) have made remarkable improvement on learning age patterns from benchmark datasets. However, for a face ``in the wild'' (from a video frame or Internet), the existing algorithms are not as accurate as for a frontal and neutral face. In addition, with the increasing number of in-the-wild aging data, the computation speed of existing deep learning platforms becomes another crucial issue. In this paper, we propose a high-efficient age estimation system  with joint optimization of age estimation algorithm and deep learning system. Cooperated with the city surveillance network, this system can provide age group analysis for intelligent demographics. First, we build a three-tier fog computing architecture including an edge, a fog and a cloud layer, which directly processes age estimation from raw videos. Second, we optimize the age estimation algorithm based on CNNs with label distribution and K-L divergence distance embedded in the fog layer and evaluate the model on the latest wild aging dataset. Experimental results demonstrate that: 1. our system collects the demographics data dynamically at far-distance without contact, and makes the city population analysis automatically; and 2. the age model training has been speed-up without losing training progress or model quality. To our best knowledge, this is the first intelligent demographics system which has potential applications in improving the efficiency of smart cities and urban living.

\end{abstract}


\section{Introduction}   

Age estimation aims to automatically predict the exact age or age group of a facial image based on the visual features.  Different from other kinds of facial information such as identity and gender, human aging is generally a slow and complicated process which makes the accurate prediction of a given facial image a challenging problem within the field of facial analysis. Facial age estimation has attracted much attention due to its potential applications in video surveillance, demographic statistics collection and business intelligence. Recently, deep learning schemes, especially Convolutional Neural Networks (CNNs), have been successfully employed for many tasks related to facial analysis including face detection, face alignment~\cite{facedetection}, face verification~\cite{deepface}, and demographic estimation~\cite{yang2011correspondence}. For facial age estimation, CNNs have been applied to learning aging features directly from large-scale aging dataset~\cite{hu2017facial} and the deeply-learned aging patterns lead to significant performance improvement on benchmark datasets~\cite{niu2016ordinal, chen2017using}.

Although a number of deep-learning based algorithms have been successfully developed for facial age estimation, we still face the challenges for real-world applications. Specifically, the challenges are from application-level to system-level:

	\begin{enumerate}
		\item \emph{Accuracy.} The performance of age estimation is not as accurate as other kinds of demographic information such as identity and gender. How to improve the estimation performance remains a challenge research problem in computer vision.
		\item \emph{Latency.} Most of the age estimation models are tested on the benchmark datasets which are mainly front-view and neutral and only have a single face. While in the real-world surveillance video frames,  there are often several faces appearing simultaneously. The age estimation will be time-consuming when the number of people in an image/frame is increasing.
		\item \emph{Model Training.} Existing distributed deep learning systems, such as Caffe \cite{jia2014caffe} and TensorFlow \cite{abadi2016tensorflow} usually take a long time to learn a convergent model due to the high communication overhead \cite{li2014communication}. 
		\item \emph{Online Prediction.} Intelligent demographics statistics requires monitoring timeliness. Because of the mobility of urban population, the age distribution is in a dynamic change pattern.  Therefore, the age estimation result should be received and updated in time.
	\end{enumerate}

Aforementioned challenges motivate us to develop an efficient age estimation system for intelligent demography based on the deep learning and fog-computing techniques. This system takes advantage of surveillance videos for the smart city project,  which implements the population investigation dynamically at far-distance and non-contact to make the city population analysis automatically. Given a clip of surveillance video, the age distribution analysis can be dynamic and real-time. Inspired by the previous work~\cite{sun2016metaflow, gao2017resource, hu2017facial}, the system is designed based on the three-tier fog computing architecture.  When a person appears in the frame, algorithms and technologies of intelligent video analytics can extract the feature of people and explore its pattern.  We optimize deep-learning based age estimation model with label distribution and K-L distance loss and evaluate the  performance on the latest aging dataset. Given sufficient demographic clues of one area, we can utilize the social statistics to analyze the dynamic demographic profile of this place.

Intelligent surveillance and demographics is a rising research topic due to the rapid development of machine learning and communication algorithms.
Some research works have dedicated  to solve the related problems. Yi \textit{et al.}~\cite{yi2016pedestrian} proposed the pedestrian behaviors model to analyze the stationary crowd group influence based on the surveillance video shot. Ling \textit{et al.}~\cite{ling2017case}  utilized multi-tiered distributed infrastructure storage to analyze traffic for intelligent transportation system.  Wahyono \textit{et al.}~\cite{filonenko2016unattended} detected stationary objects in video surveillance systems via dual background model subtraction for intelligent surveillance systems. For intelligent demographics collection, Alharbi \textit{et al.}~\cite{alharbi2016demographic} proposed a demographic group prediction mechanism from smart device users based upon the recognition of user gestures.

However, so far to our best knowledge, there is no research work to implement the facial age estimation cooperated with surveillance systems for intelligent demographics applications. Building an intelligent demographics system via city surveillance network is to improve the efficiency of services by using urban informatics and technology. The traditional demographics collection method is labor intensive and time-consuming. It can only provide the information for a certain period of time, while the demographics of a city's population is shifting over time. In this context, intelligent demographics becomes critical to the success of smart cities.

The main novelties and contributions of this work are threefold:

	\begin{enumerate}
		\item We propose a fog-computing-based intelligent demographics system to automatically collect urban population information. To our best knowledge,
		it is the first intelligent system for demographics.
		\item We implement the state-of-the-art facial age estimation algorithm for the intelligent demography system. The experiments on benchmark datasets demonstrate the effectiveness of our algorithm
		\item To improve the efficiency of the whole system, we propose a communication-efficient distributed deep learning system, which is used to train our age estimation model efficient with reduced communication overhead. 
	\end{enumerate}

The rest of this paper is organized as follows. In Section~\ref{sec:system}, we give a brief overview of the system. Then we provide a detailed description of our approach in Section~\ref{sec:offline} and Section~\ref{sec:online}. The experiments are reported in
Section~\ref{sec:exp}. Finally, we draw the conclusion of this work in Section~\ref{sec:con}.

\section{Intelligent Demographic System: An Overview}\label{sec:system}

\begin{figure*}
	\centering
	\includegraphics[scale=0.53]{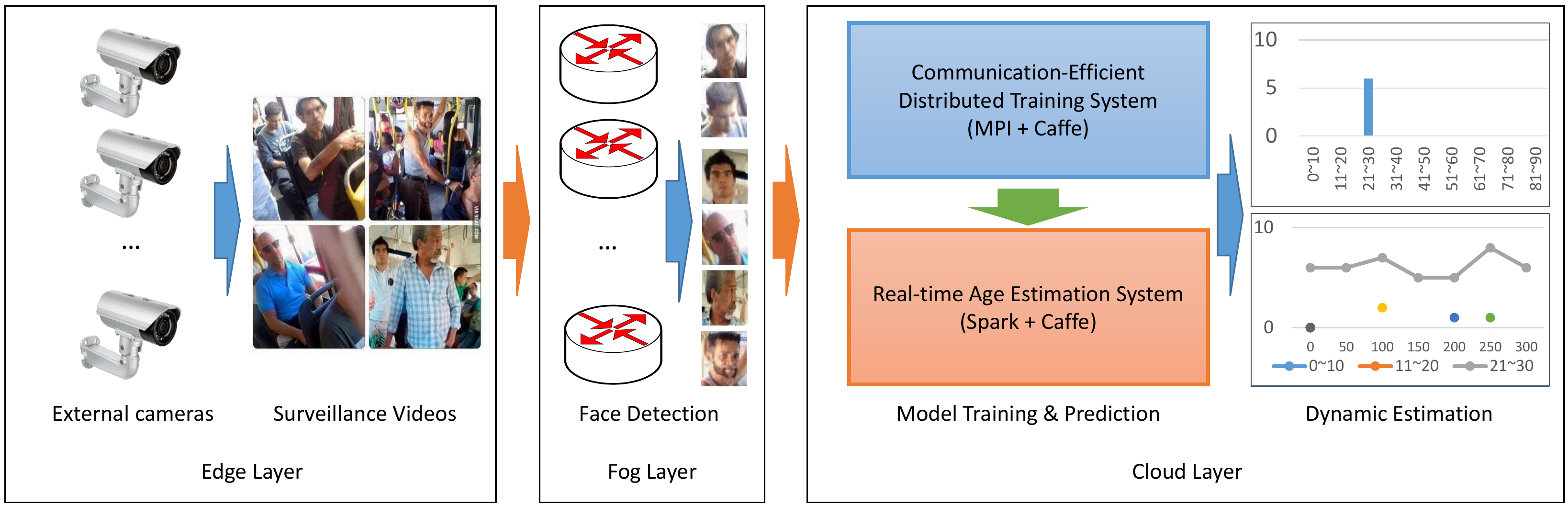}
	\caption{
		An overview of intelligent demographics system. The system is a three-tier fog computing architecture, which includes an edge, a fog and a cloud layer. 
		The edge layer captures external image data in real-time and sends them to the smart gateways. The fog layer is comprised of a set of fog nodes, like gateway, router, switch and Access Points. These smart gateways analyze all revived images, crop detected faces, and upload these face data to the cloud servers for age estimation. The cloud servers reside in top-most cloud layer, stream processing all received face data, and update demographic information.}
	\label{fig:frame}
\end{figure*}

The framework illustration of the intelligent demography system is shown in Fig.~\ref{fig:frame}. Our system is designed based on the three-tier fog computing architecture, which includes an edge, a fog and a cloud layer. Specifically, the edge layer consists of a lot of external video cameras, which are in charge of capturing image data in real-time.  The fog layer is composed of fog nodes, which usually are network devices like gateway, router, switch and Access Points. These fog nodes could collaboratively share storage and computing facilities. Traditional cloud servers reside in the top-most cloud layer, and could provide sufficient storage and computing resources. In our proposed intelligent demography system, external video cameras first send their captured images to the smart gateways. These smart gateways analyze all revived images, crop detected faces, and upload these face data to the cloud servers for age estimation. The cloud servers would stream processing all received face data, and update demographic information. 


\subsection{Edge Layer}

The edge layer indicates the external surveillance camera networks, which are the input source of the entire system. Since the surveillance system  is spread all over the city and records different scenes and events, we focus on the people domain surveillance videos and only select the cameras set in population activity areas in this paper. As the use of video surveillance cameras grows, the video resource in one city is increasing tremendously and contains redundancy visual information. In view of the general characteristics of people activities and  the computational resource, the system extracts one frame every 30 seconds and transmits it to the next layer for facial analysis.

\subsection{Fog Layer}

After the video frame capturing in the edge layer, the fog layer implements the face pre-processing, including face detection and alignment, based on the fog-computing. All the frames from the edge layer will be filtered with face detection algorithm. Face detection is carried out using the OpenCV Face Detector, which is an implementation of the Viola-Jones Face Detector~\cite{viola2001rapid} uses a boosted rejection cascade based on AdaBoost. Given the face location and area, the image of face region can be cropped and normalized. 68 face landmark
points are located by the OMRON face alignment algorithm. Faces are aligned according to the locations of two eyes and that of the mouth. The distance of two eye centers is set to 32 pixels. In a face image, the size of face bounding boxes is $128 \times 128$. The images which contain non-frontal faces are removed. 

\subsection{Cloud Layer}

This layer contains two core systems. Specifically, a distributed model training system is set up to learn a deep learning model for age estimation from the training dataset. A real-time age estimation system would leverage the learned model to process received face data, and update the demographic information accordingly.

\subsubsection{Distributed Model Training with Caffe-MPI}
We use Caffe \cite{jia2014caffe} as the computation engine for model training. The training system aims to find optimal parameters for a deep learning model to minimize its prediction error.
To achieve better performance, we leverage MPI (Message Passing Interface) to parallelize the model training in a cluster with multiple GPU nodes based on the Parameter Server (PS) framework \cite{li2014communication}. 
When using the PS framework,  Caffe-MPI evenly partitions the training dataset across multiple GPU nodes at the beginning of the training processing. During the computation, each GPU worker node fetches a batch of the assigned training data into memory,  uses an optimization method like stochastic gradient decent (SGD) to compute an update value for each parameter, and pushes the updates to a server node. When receiving updates from all GPU nodes, the server node uses these data to update the model's parameters. Next, each GPU node pulls newly computed parameters from the server node for the next iteration of the computation.  Caffe-MPI would perform the push-pull operations to update the model's parameters until convergence. It should be noted that each GPU node needs to push all updates and pull all parameters in each iteration, resulting in high communication overhead. In Section IV, we propose Caffe-ASU to address this problem.

\subsubsection{Real-time Age Estimation with Caffe-Spark}
We set up a steam processing system on the cloud layer to use the learned model to estimate received face data from the fog layer. In this system, we combine Spark Steaming \cite{zaharia2013discretized} and Caffe. Specifically, Spark Streaming is an extension of Spark  that enables scalable, high-throughput, fault-tolerant stream processing of live data streams. It structures a streaming computation as a series of stateless, deterministic batch computations on small time interval. In this system, we place the face data received every second into an interval, and run a Caffe operation on each interval to compute the age information and update the demographic information. Processed data would be stored in the local file system for future data analytics.

\section{Application Optimization: A Deep Learning Approach for Age Estimation}\label{sec:offline}
\begin{figure*}
	\centering
	\includegraphics[scale=0.5]{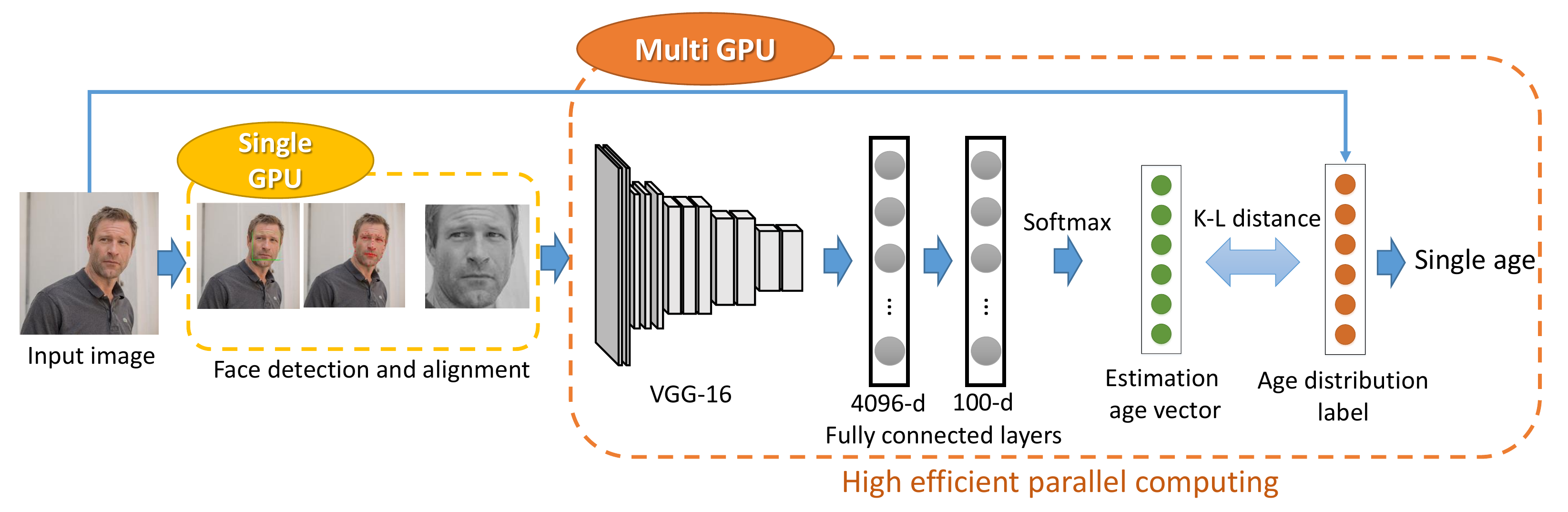}
	\caption{The facial age estimation model learning architecture.  The pre-processing of face detection and alignment is implemented on single GPU and the aligned face is the input date of CNNs.  On the top layer, we use K-L distance as the loss function. The training progress is implemented on multi-GPU with parallel computing.
	}
	\label{fig:offline}
\end{figure*}

The off-line part of the intelligent demography system is deep learning based the facial age estimation. The network architecture is shown in Fig.~\ref{fig:offline}. After the pre-processing of input face images, such as face detection and face alignment, the normalized training data are feed into Convolutional Neural Networks (CNNs). 

\subsection{Basic Network Structures}

Deep convolutional networks have witnessed great successes in computer vision area and many powerful network architectures have been developed, such as AlexNet~\cite{krizhevsky2012imagenet}, GoogLeNet~\cite{szegedy2015going}, VGGNet~\cite{simonyan2014very}, ResNet~\cite{he2016deep} and DenseNet~\cite{huang2016densely}. As the surveillance videos from one city are much larger than any existing image datasets, we need to a tradeoff between the estimation performance and computational cost when building our system.

The state-of-the-art pre-trained deep network for age estimation is DEX (Deep EXpectation of apparent age)~\cite{Rothe-ICCVW-2015}, an age estimation model based on the VGG-16 network architecture learned from the IMDB-WIKI dataset. This dataset includes more than 0.5 million images of celebrities from IMDb and Wikipedia, which is the largest publicly available dataset of face images with age labels. 

\subsection{Training Objective of Age Model}

The original pre-trained DEX model employs the softmax function in the loss layer to train the age model. However, human aging is generally a slow and smooth process in reality, and therefore cannot be treated as a
single label classification problem. In our previous work, Hu \textit{et al.}~\cite{hu2017facial} optimized the age estimation target with multi age discrete distribution vector as the ground truth label. The experimental results reflect that label distribution not only can increase the number of labeled data but also tends to learn the similarity among the neighboring ages.

In this paper, we use Gaussian distribution to model the label distribution of ages. Let $C = \{1, 2, ..., c\}$ denote the set of possible ground truth ages and $L_m = (l_m^1, l_m^2, ..., l_m^c)$ is the label distribution for the $m$-th image. Given a chronological age $a\in C$, the distribution of ages $\{a - 2, a - 1, a, a + 1, a + 2\}$ is calculated as $l_m^{a_i} = l_m^a \times e^{\frac{-(a-a_i)^2}{2\theta}} $, where the Gaussian function has the mean value $a_i$ and variance $\theta$.  For other ages, we just let $l_m^{a_i} = 0$. Finally, a normalization process is calculated  to make sure that $\sum_j^c l_m^j = 1$.

At the top layer of the deep architecture, the Kullback-Leibler (K-L) divergences distance is set to quantify the dissimilarity between the predicted label distribution to the ground truth distribution. According to the definition of K-L divergences, the distance between two discrete probability distributions $P \in R^i,Q \in R^i$ is 
\begin{equation}
\begin{aligned}
D_{KL}(P\| Q)& = \sum_{i} P_i\log \frac{P_i}{Q_i} \\
& = \sum_{i} P_i\log(P_i) - P_i\log(Q_i).
\end{aligned}            
\end{equation}
In particular, given the training data with the Gaussian label distribution, after through the shared sub-network, an image $m$ is mapped to a $c$-dimensional probability score $Q_m \in R^c$ $(Q_m^j = exp(f_m^j) / \sum_{k=1}^{c} exp(f_m^k) )$, where $f_m$ is the $c$-dimensional intermediate feature of the output of the shared sub-network for the image $m$ and $Q_m^j$ is the probability that image $m$ is in age $j$. The loss for the image $m$ is defined by
\begin{equation}
\begin{aligned}
min loss & =  \sum_{c}^{j=1} l_m^j\log(l_m^j) - l_i^j\log(Q_m^j) =  \sum_{c}^{j=1} -l_m^j\log(Q_m^j).
\end{aligned}            
\end{equation}
We optimize the network parameters via back propagation. The gradient of the softmax function is 
\begin{equation}
\frac{\partial Q_m^j}{\partial f_m^j} = Q_m^j (1 - Q_m^j).
\end{equation}
Here we provide the gradient of $loss$  with respect to $f_m^j$:
\begin{equation}
\begin{aligned}
\frac{\partial loss}{\partial f_m^j} &= \frac{\partial loss}{\partial Q_m^j} \cdot \frac{\partial Q_m^j}{\partial f_m^j} \\
&= -l_m^j \cdot \frac{1}{Q_m^j} \cdot Q_m^j(1-Q_m^j) \\
&= Q_m^j - l_m^j.
\end{aligned}            
\end{equation}

\section{System Optimization: Communication Efficient Distributed Model Training}\label{sec:online}

In this section, we propose a method to  reduce the communication overhead for the traditional PS-based distributed model training system, and implemente it into Caffe.

\begin{figure}
	\centering
	\includegraphics[scale=0.48]{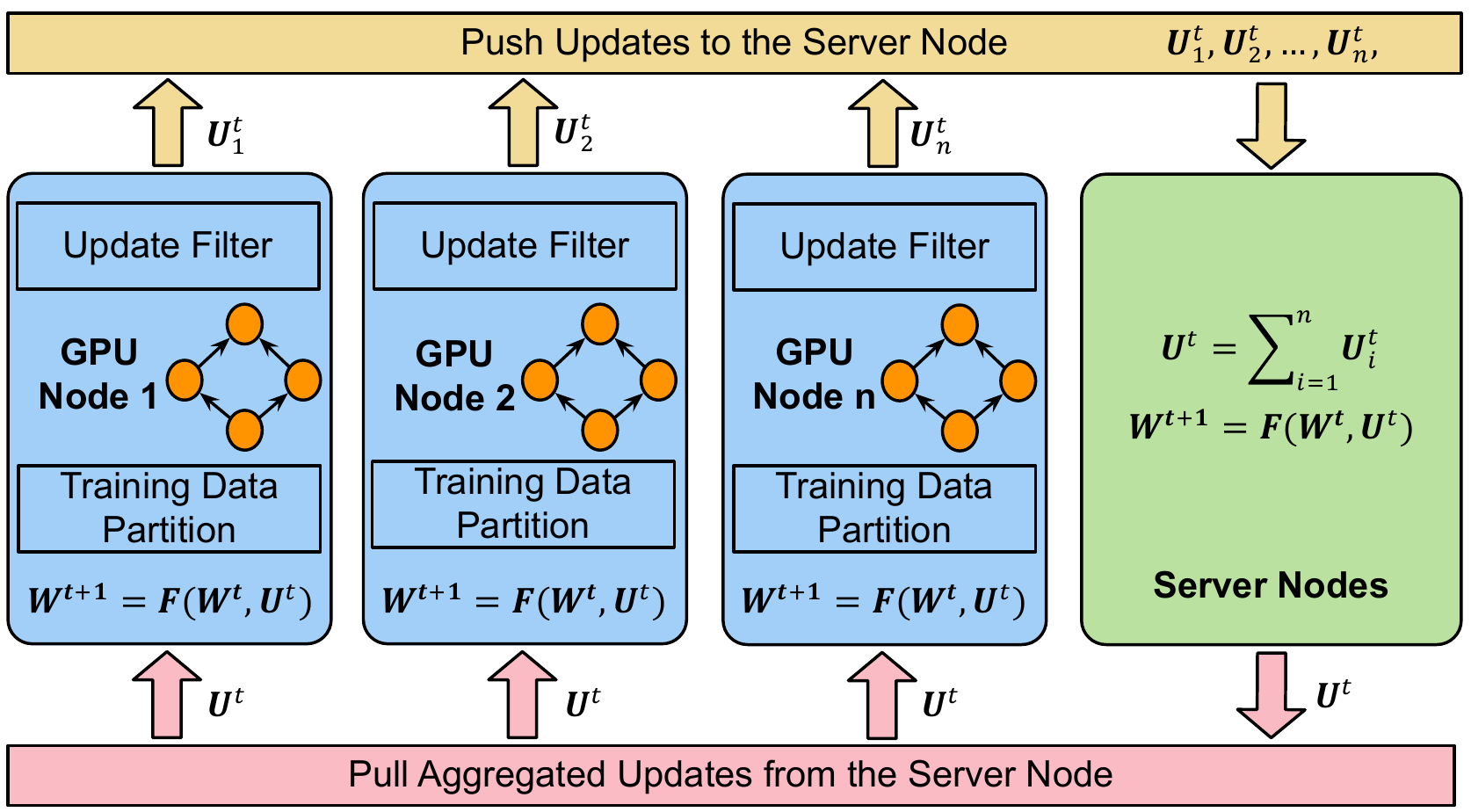}
	\caption{The system architecture of our proposed distributed model training system. Our system is designed based on the PF framework, which contains a set of GPU nodes and a server node. During the computation, each GPU node pushes generated updates $U$ to the server node for aggregation, then pulls the aggregated updates to update the model parameters $W$. To reduce network overhead, we add a filter on each GPU node.}
	\label{fig:pf}
\end{figure}

\subsubsection{Distributed Model Training with the PS Framework}

To handle large-scale DL applications, a set of distributed model training systems like TensorFlow have been proposed based on the PS framework to execute {data-parallel ML algorithms}. The PS framework contains a group of server nodes and a group of GPU worker nodes. The training model's parameters are globally shared and managed on the server nodes. Training dataset is partitioned and assigned to the GPU worker nodes. In this work, we only consider the case with multiple GPU worker nodes and one server node.

A data-parallel ML algorithm usually executes the following equation iteratively on the PS framework until some convergence criteria are met:
\begin{equation} \label{Equation: Common_Dist_ML_Alg}
\begin{alignedat}{3}
& \text{GPU Worker Nodes}: \quad && {U}^t_i  = \Delta({W}^{t}, \mathcal{D}_i), \\
& \text{Server Node}: \quad && {W}^{t+1}  = F({W}^{t}, \sum\nolimits_{i=1}^{n}{U}^t_i),
\end{alignedat}
\end{equation}
where $i$ is the index of the $i$-th GPU worker node, ${W}^{t}$ denotes the parameter vector at $t$-th iteration, ${U}^{t}_{i}$ is the update vector computed by $i$-th GPU node at $t$-th iteration using the function $\Delta(\cdot)$ and input data set $\mathcal{D}_i)$, and $F({\cdot})$ is the function used to update the model parameter vector using aggregated updates.
During the training process, a GPU node continuously performs computation on  ${W}$ and outputs ${U}$, which is aggregated on the server node to update ${W}$. To exchange data between GPU nodes and the server node, the PS framework defines a push/pull communication model: GPU worker nodes push computed ${U}$ to the \emph{server} node, and pull latest ${W}$ from it.  

\subsubsection{Caffe-ASU}

We design a filter to allow each GPU  node to selectively drop some  entries of the update vector during the push operation to reduce network traffic and communication time. In the next push operation, dropped updates would be accumulated into the newly generated update vector. In this way, each GPU worker node would push aggregated sparse updates (ASU) to the server node, rather than push all updates in each iteration. We implement this file in Caffe-MPI, and name the system Caffe-ASU.

\begin{figure}
	\centering
	\includegraphics[scale=0.73]{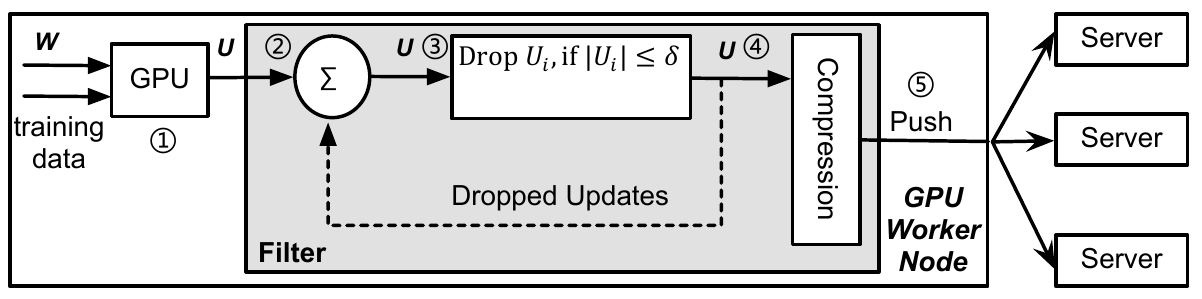}
	\caption{Caffe-ASU allows  GPU worker nodes to selectively drop {updates} with a given threshold during the push operations.}
	\label{fig:filter}
\end{figure}

As shown in Fig.~\ref{fig:filter},  before the push operations, Caffe-ASU selectively drops some {updates} as follows:
\begin{equation} 
\begin{cases}
{U}^t_{drop, i} = {U}^t_i, \; & \text{if} \; |{U}^t_i| \leq \delta\\
{U}^t_{rmn, i} = {U}^t_i,    \; & \text{if} \; |{U}^t_i| > \delta\\
\end{cases}
,
\end{equation}
where ${U}^t_{rmn}$ denotes the updates that are pushed to the server node by $i$-th GPU node, ${U}^t_{drop, i}$ contains all dropped updates in this push operation, and $\delta$ is a predefined threshold. In the next communication operation, ${U}^t_{drop}$ would be accumulated into the newly generated update vector as
\begin{equation}
{U}^{t+1}_i \gets {U}^t_{drop, i} + {U}^{t+1}_i.
\end{equation}

Since each GPU node just sends a partition of updates to the server node, Caffe-ASU would not update all parameters in each iteration. Therefore, during the pull operation, GPU worker nodes only  pull updated parameters. Our experiments (see Section V)  showed that this policy could reduce communication time by a factor of 30 without losing training progress or model quality.

\section{Experiments and Numerical Results}\label{sec:exp}
In this section, we evaluate the efficiency and effectiveness of proposed intelligent demography system. The following describes the details of the experiments and results.

\subsection{Age Model Performance Evaluation}\label{sec:age-database}

\subsubsection{Age Training Datasets}
To make the age model more appropriate for the real world application,  we set our experiments based on two latest in-the-wild  aging datasets crawled from web, i.e. IMDB-WIKI and AgeDB dataset. The details are listed in Table~\ref{tab:tab1}.

\textbf{IMDB-WIKI dataset}~\cite{Rothe-ICCVW-2015}  is the largest publicly available  dataset for age estimation of people in the wild containing 460,723 face images from 20,284 celebrities from IMDb and 62,328 from Wikipedia, thus 523,051 in total. According to the query list including the most popular 100,000 actors on the IMDb website, the profiles date of birth, name, gender and all images related to actors are crawled. 

\textbf{AgeDB dataset}~\cite{moschoglou2017agedb} is an in-the-wild
dataset with large variations in pose, expression and illuminations.  It contains $16,488$ images of various famous people with accurate to the year, noise-free labels.  Every image is annotated with respect to the identity, age and gender attribute. There exist a total of 568 distinct subjects. The average number of images per subject is 29. The minimum and maximum age are 1 and 101, respectively.

\begin{table}[htbp]
\caption{An overview of the used dataset for off-line age model training}
\begin{center}
\begin{tabular}{|c|c|c|c|}
\hline
\textbf{Dataset} & \textbf{\# Images}& \textbf{\# Subjects}& \textbf{\# Year} \\\hline
IMDB-WIKI~\cite{Rothe-ICCVW-2015} & 523,051 & 20,284 & 2015 \\\hline
AgeDB~\cite{moschoglou2017agedb} & 16,488 & 568 & 2017 \\\hline
\end{tabular}
\label{tab:tab1}
\end{center}
\end{table}

\subsubsection{Age Estimation Evaluation}

To evaluate the performance of age estimation algorithm, we use the Mean Absolute Error (MAE) as the evaluation measures The MAE is calculated based on the average of the absolute errors between the estimated age and the ground truth (labeled age), which is represented as
\begin{equation}
MAE = \frac{1}{N}\sum_{n=1}^{N}\|l_n-y_n\|,
\end{equation}
where $l_n$ is the ground truth label of the $n$th image and $y_n$ represents the estimated age based on the proposed framework. $N$ is the total number of testing samples. 

And in this work, we also consider another measurement specifically for the demography estimation.  We evaluate the estimation accuracy of age group with different age gaps. The goal is to predict whether a person's age within some range instead of  predicting the precise biological age. Because in demography analysis, the age structure of a population refers to the number of people in different age groups. Here we evaluate the performance of age group with different age ranges: 5 years, 10 years, 15 years and 20 years.

\subsubsection{Age Estimation Performance}
\begin{figure}[t!]
	\centering
	\subfigure[AgeDB]{\includegraphics[width=4.3cm]{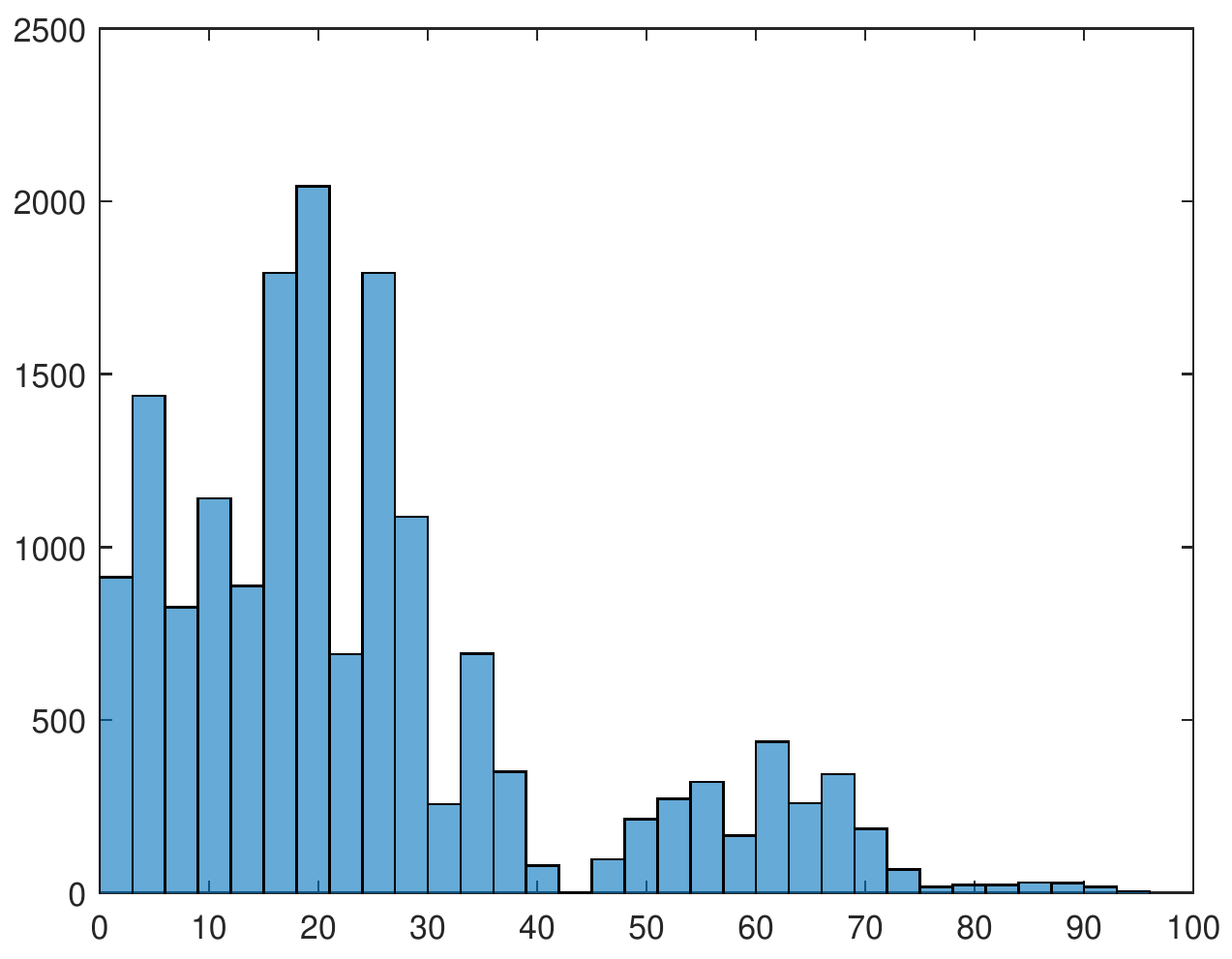}\label{fig:sub1}}
	\subfigure[IMDB-WIKI]{\includegraphics[width=4.3cm]{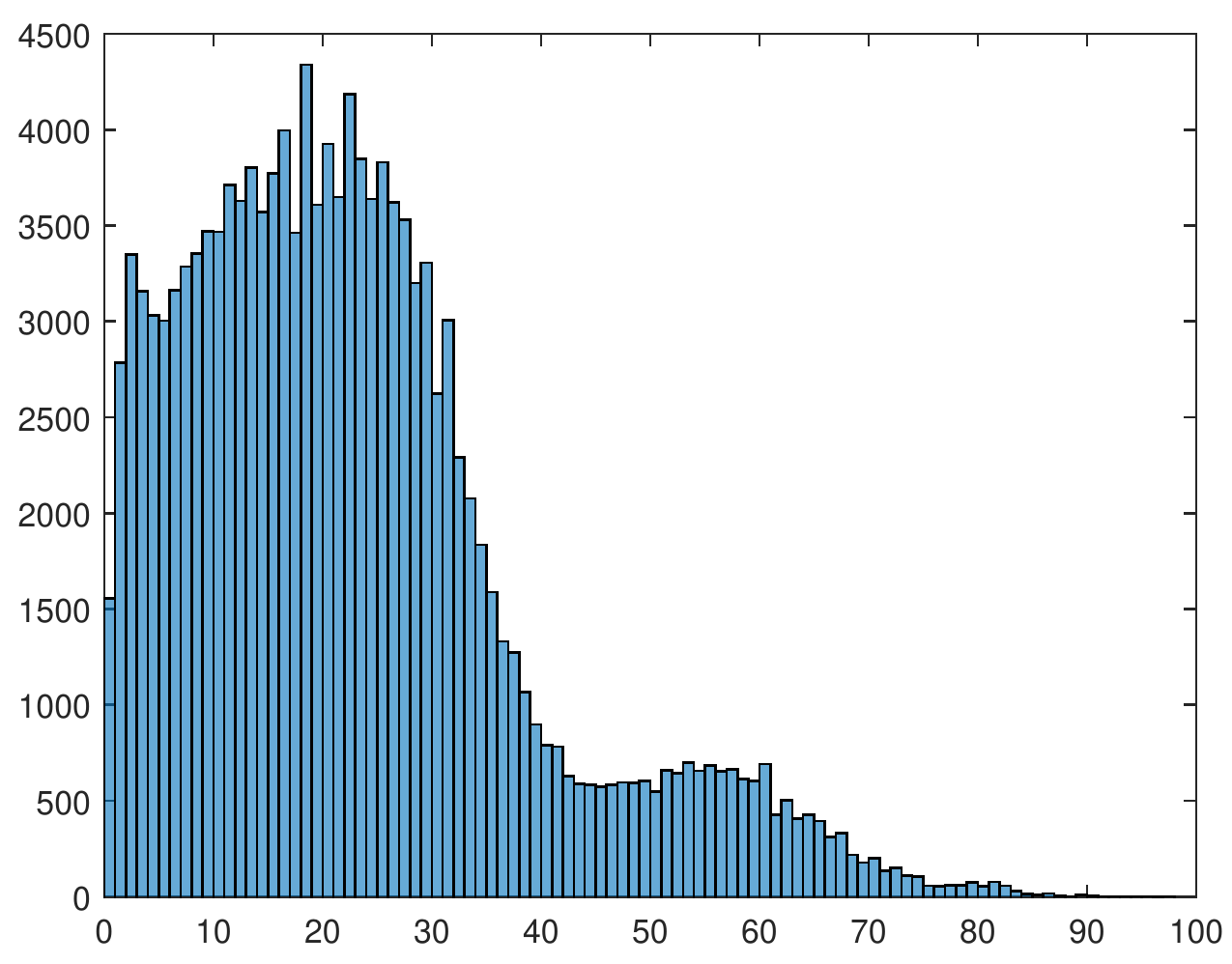}\label{fig:sub2}} \\
	\caption{The MAE statistical results of age model testing on AgeDB and IMDB-WIKI dataset.}\label{fig:result}
\end{figure}

We first test the DEX model, which is pre-trained on the IMDB-WIKI dataset with softmax loss, on the whole AgeDB dataset. The MAE of estimation results is 28.47. Then we fine tune the whole network based on the AgeDB dataset by replacing the loss layer with K-L divergences distance and the MAE has dropped into 20.32. We also test the fine-tuned age estimation model on the IMDB-WIKI dataset and the MAE is reduced from 38.4 to 22.5. The detailed MAE comparison is listed in Tabel~\ref{tab:mae}. To be more clear illustration, we summarize the MAE distribution on the two aging datasets and show the result in Figure~\ref{fig:result}.

\begin{table}
	\caption{MAE comparison on AgeDB and IMDB-WIKI datasets.}
	\begin{center}
	\begin{tabular}{c|>{\centering\arraybackslash}p{2.8cm}|>{\centering\arraybackslash}p{2.8cm}}
		\hline
		\multirow{2}{*}{\textbf{Dataset}} &
		\multicolumn{2}{c}{\textbf{Model}} \\
		\cline{2-3} 
		& {DEX~\cite{Rothe-ICCVW-2015}} & {Fine-tuned Model} \\
		\midrule
		IMDB-WIKI & 38.4 & \textbf{22.5} \\\hline
		AgeDB & 28.47 & \textbf{20.32} \\\hline
	\end{tabular}
	\label{tab:mae}
	\end{center}
\end{table}

From this result we can see that the multi-age distribution label and the K-L distance loss can improve the wild age estimation significantly. Since the age estimation task for image in the wild is very challenging, We also list the age group accuracy in Table~\ref{tab:tab2}.

\begin{table}[htbp]
	\caption{The age group estimation of different age ranges.}
	\begin{center}
		\begin{tabular}{>{\centering\arraybackslash}p{3cm}|>{\centering\arraybackslash}p{3cm}}
			\hline
			\textbf{Age Gap} & \textbf{Accuracy}\\\hline
			5 years & 23.6\% \\
			10 years & 43.9\% \\
			15 years & 62.3\%  \\
			20 years & 73.7\% \\\hline
		\end{tabular}
		\label{tab:tab2}
	\end{center}
\end{table}

\subsection{Model Training Efficiency Evaluation}

In this set of experiments, we measure  the performance of Caffe-ASU. We use AgeDB as the training and test dataset, and run the experiments on 4 GPU virtual machines. In addition, we also implement Caffe-DSU based on \cite{li2014communication}, which uses a threshold to drop updates for pushing without aggregating dropped one. Caffe-RAW would push all updates and pull all parameters at each iteration. In the experiments, we use $1\times10^{-5}$ as the threshold, and users could select different values according to their particular applications.

\setlength{\minipagewidth}{0.24\textwidth}
\setlength{\figurewidthFour}{\minipagewidth}
\begin{figure} 
    \centering
    \begin{minipage}[t]{\minipagewidth}
    \begin{center}
    \includegraphics[width=\figurewidthFour]{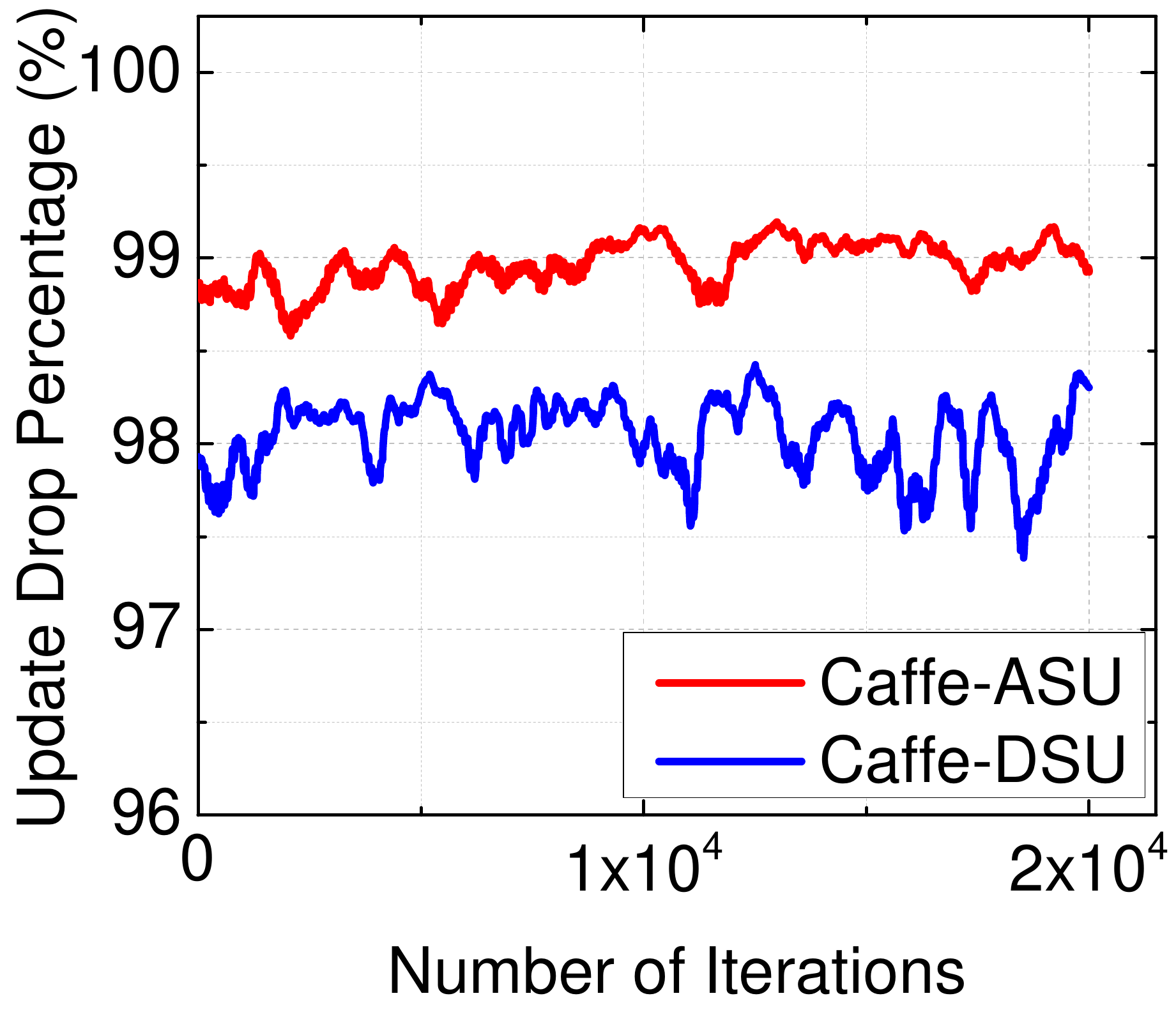}
    \subcaption{(a) Percentage of Dropped Updates}
    \end{center}
    \end{minipage}
    \centering
    \begin{minipage}[t]{\minipagewidth}
    \begin{center}
    \includegraphics[width=\figurewidthFour]{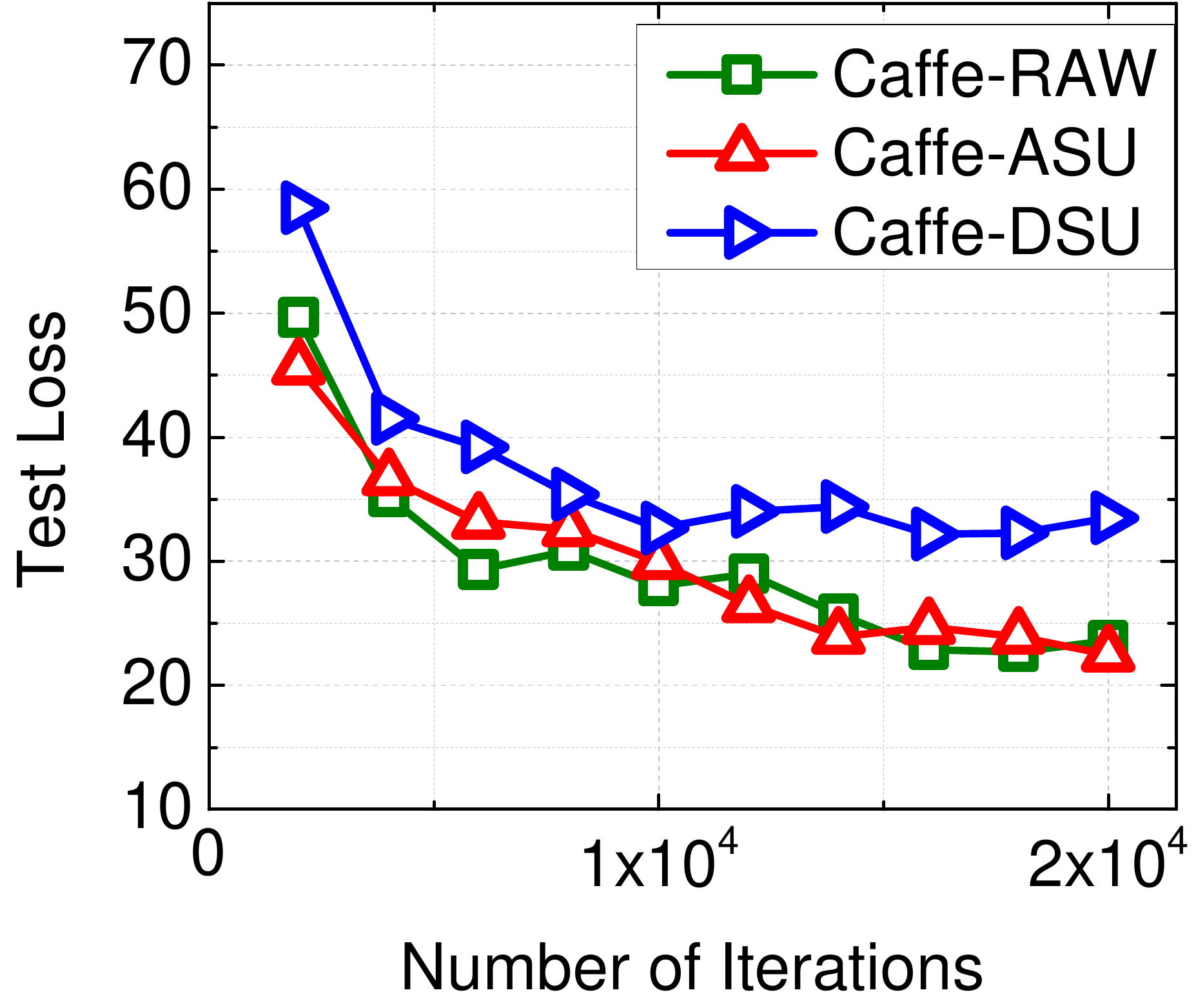}
    \subcaption{(b)  Test Loss}
    \end{center}
    \end{minipage}
    \centering
    \begin{minipage}[t]{\minipagewidth}
    \begin{center}
    \includegraphics[width=\figurewidthFour]{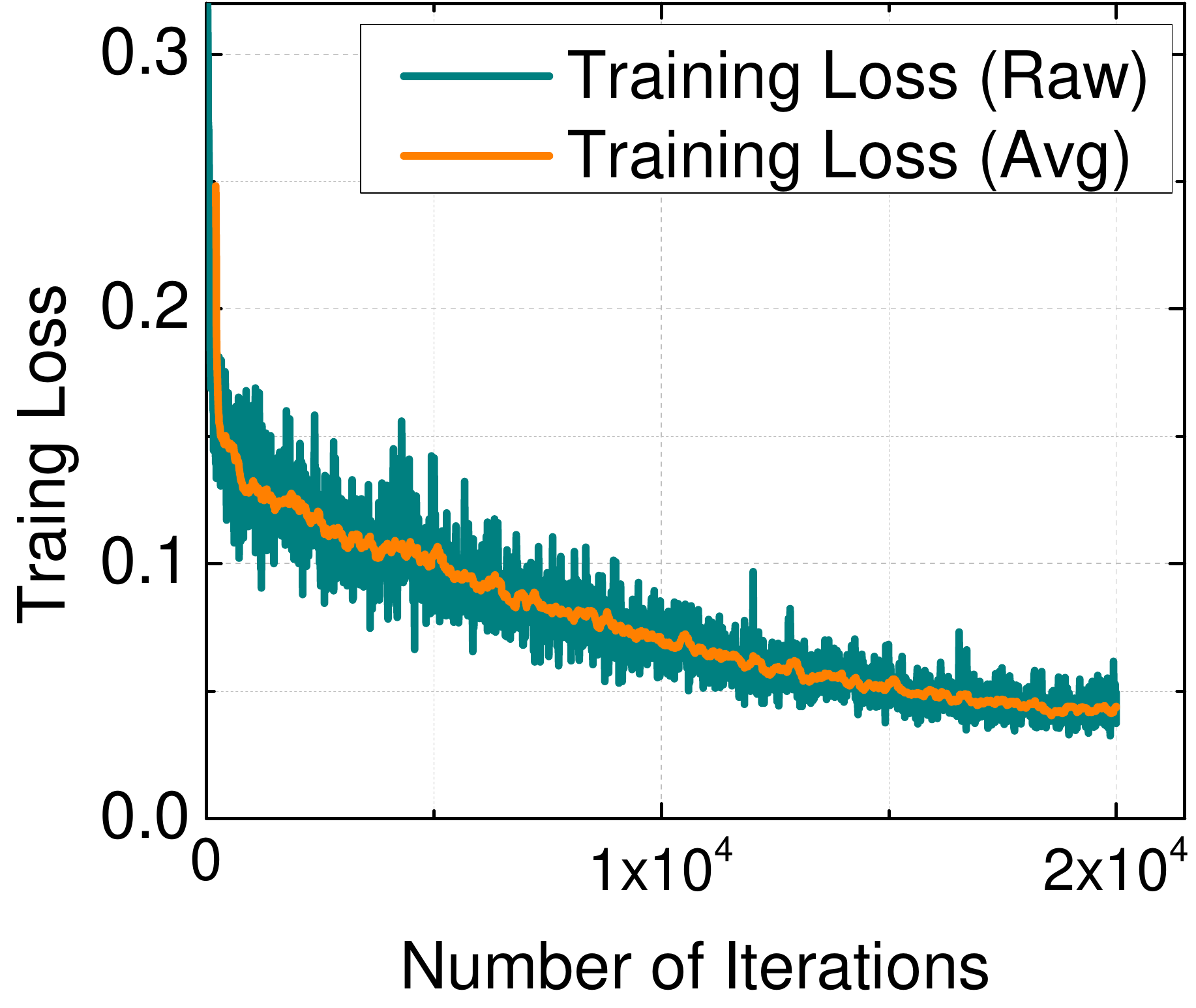}
    \subcaption{(c)  Training Loss with Caffe-RAW}
    \end{center}
    \end{minipage}
    \centering
    \begin{minipage}[t]{\minipagewidth}
    \begin{center}
    \includegraphics[width=\figurewidthFour]{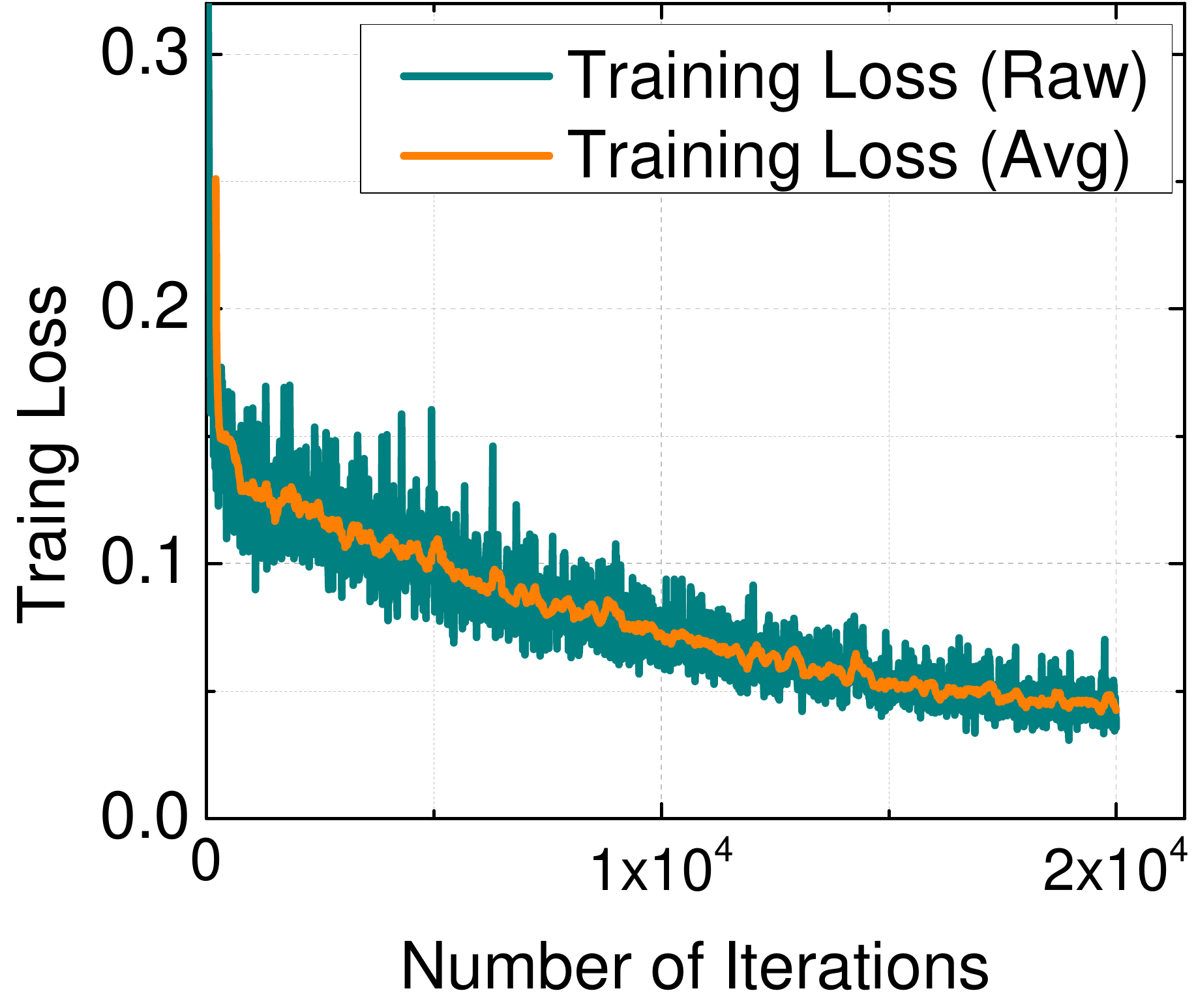}
    \subcaption{(d)  Training Loss with Caffe-ASU}
    \end{center}
    \end{minipage}
    \centering
    \caption{Performance of Caffe-ASU. (a) shows that percentage of dropped updates during the training progress. (b) shows the loss value on the test dataset. (c) (d) show the loss value on the training dataset using Caffe-RAW and Caffe-ASU, respectively.}
\label{fig:pf-loss-net}
\end{figure}

\subsubsection{Percentage of Dropped Updates}
As shown in Fig.~\ref{fig:pf-loss-net}(a), when using $1\times10^{-5}$ as the threshold, Caffe-ASU could drop about $98.8\%$ of updates at each iteration. It means that Caffe-ASU pushes about $1.2\%$ of updates to the server node, and pulls roughly $1.2\%$ of updated parameters from the server node. In addition, our proposed deep learning model contains about $135$ millions of parameters.  When using Caffe-RAW, each GPU node needs to push about $540$MB updates and pull $540$MB parameters via network. As a comparison, Caffe-ASU allows each GPU node to push only $11$MB and pull $15$MB data on average in our testbed. We also note that Caffe-DSU could drop about $98\%$ of updates at each iteration at the cost of losing model quality from Fig.~\ref{fig:pf-loss-net}(b).

\subsubsection{Model Quality}

Fig.~\ref{fig:pf-loss-net}(b) shows that Caffe-ASU would not reduce model quality, which is measured by the loss value on the testing dataset. Specifically, Caffe-ASU and Caffe-RAW could achieve the loss value of about $22.5$ on the testing dataset. We can also find that Caffe-DSU could only achieve the loss value of about $32.2$ on the testing dataset. It means that Caffe-ASU could drop about $98.8\%$ of updates without losing model quality, which Caffe-DSU could not guarantee model quality if dropping about $98\%$ of updates at each iteration. Thus, we could conclude that Caffe-ASU could achieve significant gains as compared to Caffe-DSU.

\subsubsection{Training Progress}

Fig.~\ref{fig:pf-loss-net}(b) shows that Caffe-ASU would not affect the training progress, which is measured by the loss value on the training dataset at different iterations. As we can see, at iteration $2 \times 10^{4}$, both Caffe-ASU and Caffe-RAW could achieve the loss value of about $0.038$. It means that Caffe-ASU does not meed additional iterations to achieve the same loss value on the training dataset, compared with Caffe-RAW. 

\subsubsection{Speedup Ratio} 
We run the experiments using 1Gbps and 10Gbps network, and show the computation and communication time in Fig.~\ref{fig:pf-time}. When using 1Gbps, Caffe-ASU could reduce the communication time by a factor of about 28 compared to Caffe-RAW, and speed up the training progress by a factor of about $4.9$. When using 10Gbps, Caffe-ASU could reduce the communication time by a factor of about 16 compared to Caffe-RAW, and speed up the training progress by a factor of about $1.6$. 

\setlength{\minipagewidth}{0.24\textwidth}
\setlength{\figurewidthFour}{\minipagewidth}
\begin{figure} 
    \centering
    \begin{minipage}[t]{\minipagewidth}
    \begin{center}
    \includegraphics[width=\figurewidthFour]{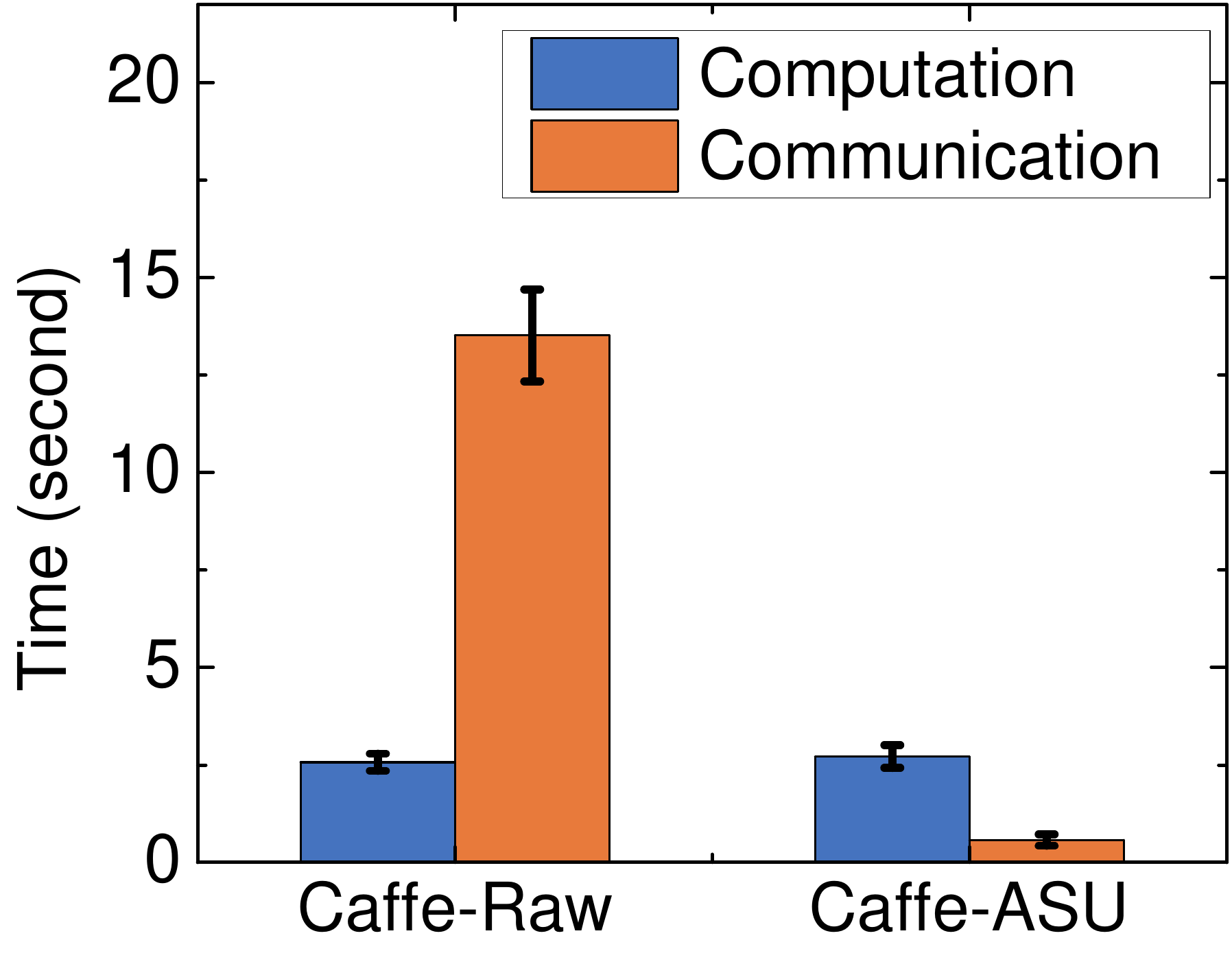}
    \subcaption{(a) 1Gbps Network}
    \end{center}
    \end{minipage}
    \centering
    \begin{minipage}[t]{\minipagewidth}
    \begin{center}
    \includegraphics[width=\figurewidthFour]{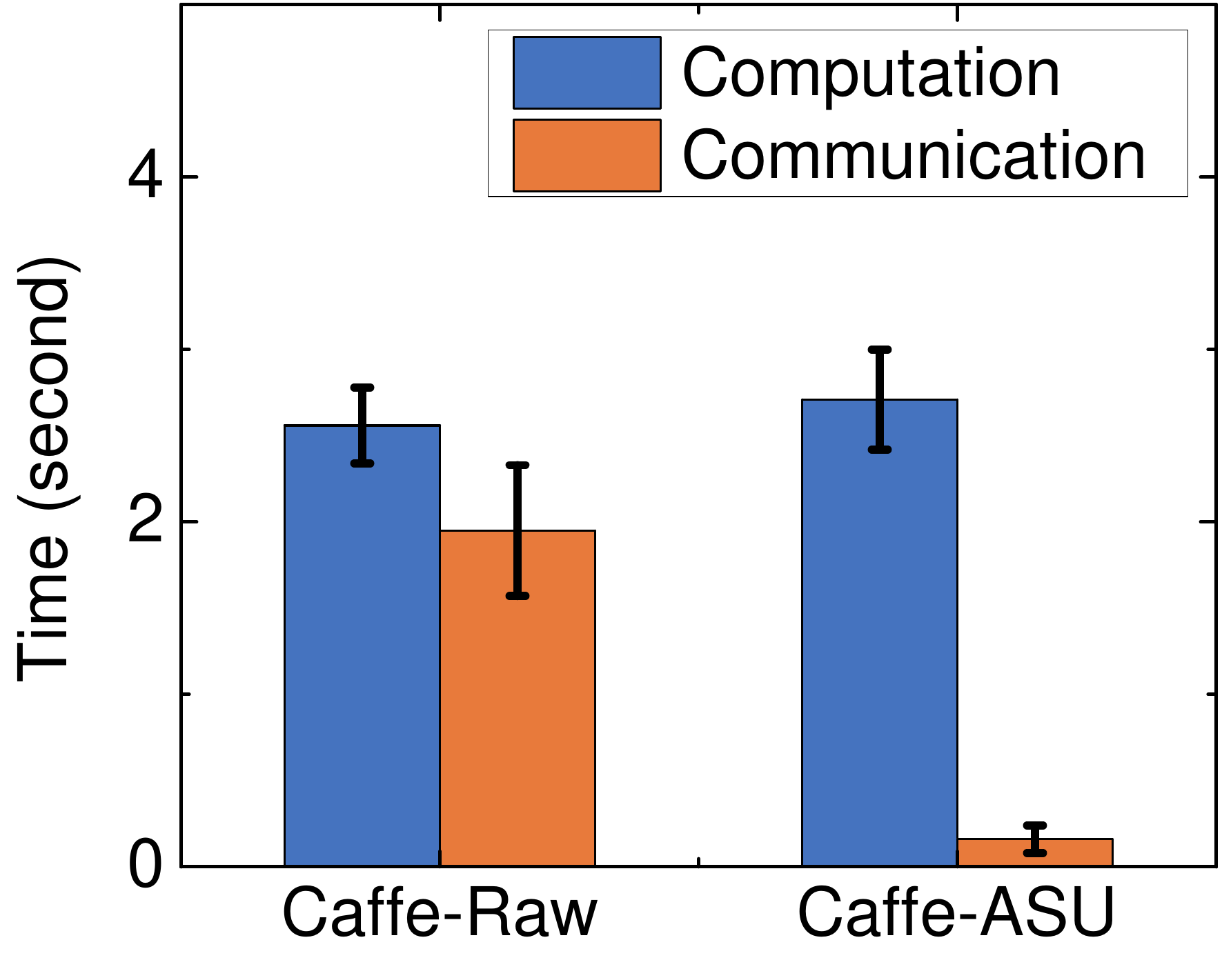}
    \subcaption{(b) 10Gbps Network}
    \end{center}
    \end{minipage}
    \centering
    \caption{Speedup ratio of Caffe-ASU using 1Gbps and 10Gbps network.}
\label{fig:pf-time}
\end{figure}

\section{Conclusion}\label{sec:con}

In this paper, we investigate the problem of speeding-up facial age estimation. We propose a high-efficient age estimation system with joint optimization of age estimation algorithm and deep learning system. The system is designed based on the three-tier fog computing architecture and provides the age group analysis directly from raw videos. Then we apply the system for intelligence demographics. Experimental results demonstrate the effectiveness and efficiency of our system. To our best knowledge, this is the first intelligent demographics system which implements the population investigation automatically via surveillance videos. In the future, we aim to further improve the performance of age estiamtion algorithm and apply the proposed system in the large-scale video surveillance of smart city project. The intelligent demographics system will improve the efficiency of smart cities and urban living.

\section{Acknowledgments}
This article was supported in part by Grant of AcRF Tier 1 RG26/16 and Tier 2 ARC 42/13 from  Singapore MOE.

\bibliographystyle{IEEEtran}
\bibliography{iccbib}
%

\end{document}